\documentclass[10pt, conference, letter, oneside, twocolumn]{IEEEtran}
\usepackage{graphicx}
\usepackage{amsmath}
\usepackage{amssymb}
\usepackage{mathrsfs}
\usepackage{stfloats}
\usepackage{dsfont}
\usepackage{setspace}
\newtheorem{corollary}{Corollary}
\newtheorem{proposition}{Proposition}
\newtheorem{lemma}{Lemma}

\begin{document}
\title{Stochastic Geometry Modeling and Performance Evaluation of mmWave Cellular Communications}
\author{\authorblockN{Marco Di Renzo}
\authorblockA{Laboratory of Signals and Systems (L2S - UMR 8506) \\
French National Center for Scientific Research (CNRS) \\
\'Ecole Sup\'erieure d'\'Electricit\'e (SUPELEC) \\
University of Paris-Sud XI (UPS-XI) \\
3 rue Joliot-Curie, 91192 Gif-sur-Yvette (Paris), France \\
E-Mail: marco.direnzo@lss.supelec.fr} }
\maketitle
\begin{abstract}
In this paper, a new mathematical framework to the analysis of millimeter wave cellular networks is introduced. Its peculiarity lies in considering realistic path-loss and blockage models, which are derived from experimental data recently reported in the literature. The path-loss model accounts for different distributions for line-of-sight and non-line-of-sight propagation conditions and the blockage model includes an outage state that provides a better representation of the outage possibilities of millimeter wave communications. By modeling the locations of the base stations as points of a Poisson point process and by relying upon a noise-limited approximation for typical millimeter wave network deployments, exact integral expressions for computing the coverage probability and the average rate are obtained. With the aid of Monte Carlo simulations, the noise-limited approximation is shown to be sufficiently accurate for typical network densities. Furthermore, it is shown that sufficiently dense millimeter wave cellular networks are capable of outperforming micro wave cellular networks, both in terms of coverage probability and average rate.
\end{abstract}
\section{Introduction} \label{Introduction}
\PARstart{I}{n} spite of common belief, recently conducted channel measurements have shown that millimeter wave (mmWave) frequencies may be suitable for cellular communications, provided that the cell radius is of the order of 100-200 meters \cite{Rappaport__mmWaveAccess}. Based on these measurements, the authors of \cite{Rappaport__mmWaveJSAC} have recently studied system-level performance of mmWave cellular networks and have compared them against conventional micro wave ($\mu$Wave) cellular networks. The results have highlighted that mmWave cellular communications may outperform $\mu$Wave cellular communications, by assuming similar cellular network densities, provided that a sufficient beamforming gain is guaranteed between Base Stations (BSs) and Mobile Terminals (MTs). These preliminary but encouraging results have motivated several researchers to investigate potential and challenges of mmWave cellular communications for wireless access, in light of the large and unused bandwidth that is available at these frequencies.

System-level performance evaluation of cellular networks is widely recognized to be a mathematically intractable problem \cite{AndrewsNov2011}, due to the lack of tractable approaches for modeling the other-cell interference. Only recently, a new mathematical methodology has gained prominence due to its analytical tractability and reasonable accuracy compared to currently deployed cellular networks. This emerging approach exploits results from stochastic geometry and relies upon modeling the locations of the BSs as points of a point process \cite{AndrewsNov2011}. Usually, the Poisson Point Process (PPP) is used due to its mathematical tractability \cite{Haenggi_Survey2013}. Recent results for cellular networks modeling based on stochastic geometry are available in \cite{MDR_TCOMrate}-\cite{MDRwei_TCOM2015}, to which the reader is referred for further information.

Motivated by the mathematical flexibility of the PPP-based abstraction modeling, researchers have recently tried to study the performance of mmWave cellular communications with the aid of stochastic geometry, by developing new frameworks specifically tailored to account for the peculiarities of mmWave propagation \cite{Heath__mmWave}, \cite{Singh__mmWave}. In fact, the mathematical frameworks currently available for modeling $\mu$Wave cellular networks are not directly applicable to mmWave cellular networks. The main reasons are related to the need of incorporating realistic path-loss and blockage models, which are significantly different from $\mu$Wave communications. For example, \cite{Rappaport__mmWaveAccess} and \cite{Rappaport__mmWaveJSAC} have pointed out that Line-Of-Sight (LOS) and Non-Line-Of-Sight (NLOS) links need to be appropriately modeled and may have different distributions, due to the more prominent impact of spatial blockages at mmWave frequencies compared to $\mu$Wave frequencies. Furthermore, in mmWave communications a new outage state may be present in addition to LOS and NLOS states, which better reflects blockage effects at high frequencies and accounts for the fact the a link may be too weak to be established.

Recently reported results on stochastic geometry modeling of mmWave cellular communications \cite{Heath__mmWave}, \cite{Singh__mmWave} take these aspects into account only in part. The approach proposed in \cite{Heath__mmWave} is mathematically tractable for dense network deployments and relies upon an equivalent LOS ball approximation. The approach proposed in \cite{Singh__mmWave} uses a similar LOS ball approximation, but is applicable to medium/sparse network deployments. The outage state observed in \cite{Rappaport__mmWaveJSAC} is not explicitly taken into account either in \cite{Heath__mmWave} or in \cite{Singh__mmWave}. Against this background, in the present paper a new methodology to the stochastic geometry modeling of mmWave cellular communications is proposed, which explicitly incorporates the empirical path-loss and blockage models recently reported in \cite{Rappaport__mmWaveJSAC}.

The paper is organized as follows. In Section \ref{SystemModel}, system model and main modeling assumptions are introduced. In Section \ref{CoverageRate}, the mathematical frameworks for computing coverage probability and average rate are described. In Section \ref{Results}, the frameworks are validated via numerical simulations and the performance of mmWave and $\mu$Wave cellular networks are compared. Finally, Section \ref{Conclusion} concludes this paper.
\section{System Model} \label{SystemModel}
\subsection{PPP-Based Abstraction Modeling} \label{PPP_CellularModeling}
A bi-dimensional downlink cellular network deployment is considered, where a probe MT is located, without loss of generality thanks to the Slivnyak theorem \cite[vol. 1, Th. 1.4.5]{BaccelliBook2009}, at the origin and the BSs are modeled as points of a homogeneous PPP, denoted by $\Psi $, of density $\lambda$. The MT is assumed to be served by the BS providing the smallest path-loss to it. The path-loss model is introduced in Section \ref{PathLossModeling}. The serving BS is denoted by ${\rm{BS}}^{\left( 0 \right)}$. Similar to \cite[Sec. VI]{AndrewsNov2011}, full-frequency reuse is considered. For notational simplicity, the set of interfering BSs is denoted by $\Psi ^{\left( \backslash 0 \right)}  = \Psi \backslash {\rm{BS}}^{\left( 0 \right)}$. The distance from the generic BS to the MT is denoted by $r$.
\subsection{Directional Beamforming Modeling} \label{BeamformingModeling}
Thanks to the small wavelength, mmWave cellular networks are capable of exploiting directional beamforming for compensating for the increased path-loss at mmWave frequencies and for overcoming the additional noise due to the large transmission bandwidth. As a desirable bonus, directional beamforming provides interference isolation, which reduces the impact of the other-cell interference. Thus, antenna arrays are assumed at both the BSs and the MT for performing directional beamforming. For mathematical tractability and similar to \cite{Heath__mmWave} and \cite{Singh__mmWave}, the actual antenna array patterns are approximated by considering a sectored antenna model. In particular, the antenna gain of the generic BS, $G_{{\rm{BS}}} \left(  \cdot  \right)$, and of the MT, $G_{{\rm{MT}}} \left(  \cdot  \right)$, can be formulated as follows:
\setcounter{equation}{0}
\begin{equation}
\label{Eq_1}
G_{\rm{q}} \left( \theta  \right) = \begin{cases}
 G_{\rm{q}}^{\left( {\max } \right)} & {\rm{if}}\quad \left| \theta  \right| \le \omega _{\rm{q}}  \\
 G_{\rm{q}}^{\left( {\min } \right)} & {\rm{if}}\quad \left| \theta  \right| > \omega _{\rm{q}}  \\
 \end{cases}
\end{equation}
\noindent where $q \in \left\{ {{\rm{BS}},{\rm{MT}}} \right\}$, $\theta  \in \left[ {0,2\pi } \right)$ denotes the angle off the boresight direction, $\omega _q $ denotes the beamwidth of the main lobe, $G_q^{\left( {\max } \right)}$ is the array gain of the main lobe and $G_q^{\left( {\min } \right)}$ is the antenna gain of the side lobe.

Both the MT and its serving BS, ${\rm{BS}}^{\left( 0 \right)}$, are assumed to estimate the angles of arrival and to adjust their antenna steering orientations accordingly. On the intended links, therefore, the maximum directivity gain can be exploited. Thus, the directivity gain of the intended link is $G^{\left( 0 \right)}  = G_{{\rm{BS}}}^{\left( {\max } \right)} G_{{\rm{MT}}}^{\left( {\max } \right)}$. The beams of all non-intended links are assumed to be randomly oriented with respect to each other and to be uniformly distributed in $\left[ {0,2\pi } \right)$. Accordingly, the directivity gains of the interfering links, $G^{\left( i \right)}$ for $i \in \Psi ^{\left(\backslash 0 \right)}$, are randomly distributed. Based on \eqref{Eq_1}, their Probability Density Function (PDF) can be formulated as $f_{G^{\left( i \right)} } \left( g \right) = \left( {f_{G_{{\rm{BS}}}^{\left( i \right)} }  \otimes f_{G_{{\rm{MT}}}^{\left( i \right)} } } \right)\left( g \right)$, where:
\setcounter{equation}{1}
\begin{equation}
\label{Eq_2}
f_{G^{\left( i \right)}_q } \left( g \right) = \frac{{\omega _{\rm{q}} }}{{2\pi }}\delta \left( {g - G_q^{\left( {\max } \right)} } \right) + \left( {1 - \frac{{\omega _{\rm{q}} }}{{2\pi }}} \right)\delta \left( {g - G_q^{\left( {\min } \right)} } \right)
\end{equation}
\noindent for $q \in \left\{ {{\rm{BS}},{\rm{MT}}} \right\}$, $\delta \left(  \cdot  \right)$ is the Kronecker's delta function and $\otimes$ denotes the convolution operator.
\subsection{Link State Modeling} \label{LinkStateModeling}
Let an arbitrary link of length $r$, \textit{i.e.}, the distance from the generic BS to the MT is equal to $r$. Motivated by recent experimental findings on mmWave channel modeling \cite[Sec. III-D]{Rappaport__mmWaveJSAC}, a three-state statistical model for each link is assumed, according to which each link can be in a LOS, NLOS or in an outage (OUT) state. A LOS state occurs if there is no blockage between BS and MT. A NLOS state, on the other hand, occurs if the BS-to-MT link is blocked. An outage state occurs if the path-loss between BS and MT is so high that no link between them can be established. In this case, the path-loss of the link is assumed to be infinite. In practice, outages occur implicitly when the path-loss in either a LOS or a NLOS state are sufficiently large. In \cite[Fig. 7]{Rappaport__mmWaveJSAC}, with the aid of experimental results, it is proved that adding an outage state, which usually is not observed for transmission at $\mu$Wave frequencies, provides a more accurate statistical description of the inherent coverage possibilities at mmWave frequencies.

In particular, the probabilities of occurrence $p_{{\rm{LOS}}} \left(  \cdot  \right)$, $p_{{\rm{NLOS}}} \left(  \cdot  \right)$, $p_{{\rm{OUT}}} \left(  \cdot  \right)$ of LOS, NLOS and outage states, respectively, as a function of the distance $r$ can be formulated, based on experimental results, as follows \cite[Eq. 8]{Rappaport__mmWaveJSAC}:
\setcounter{equation}{2}
\begin{equation}
\label{Eq_3}
\begin{array}{l}
 p_{{\rm{OUT}}} \left( r \right) = \max \left\{ {0,1 - \gamma _{{\rm{OUT}}} e^{ - \delta _{{\rm{OUT}}} r} } \right\} \\
 p_{{\rm{LOS}}} \left( r \right) = \left( {1 - p_{{\rm{OUT}}} \left( r \right)} \right)\gamma _{{\rm{LOS}}} e^{ - \delta _{{\rm{LOS}}} r}  \\
 p_{{\rm{NLOS}}} \left( r \right) = \left( {1 - p_{{\rm{OUT}}} \left( r \right)} \right)\left( {1 - \gamma _{{\rm{LOS}}} e^{ - \delta _{{\rm{LOS}}} r} } \right)
 \end{array}
\end{equation}
\noindent where $\left( {\delta _{{\rm{LOS}}} ,\gamma _{{\rm{LOS}}} } \right)$ and $\left( {\delta _{{\rm{OUT}}} ,\gamma _{{\rm{OUT}}} } \right)$ are parameters that depend on the propagation scenario and on the carrier frequency being considered. Examples are available in \cite[Table I]{Rappaport__mmWaveJSAC}.

Under the assumptions that the BSs are modeled as points of a homogeneous PPP and that the probabilities of the BS-to-MT links being in a LOS, NLOS or outage state are independent, $\Psi$ can be partitioned into three (one for each link state) independent and non-homogeneous PPPs, \textit{i.e.}, $\Psi _{{\rm{LOS}}}$, $\Psi _{{\rm{NLOS}}}$ and $\Psi _{{\rm{OUT}}}$, such that $\Psi  = \Psi _{{\rm{LOS}}}  \cup \Psi _{{\rm{NLOS}}}  \cup \Psi _{{\rm{OUT}}}$. This originates from the thinning property of the PPPs \cite{BaccelliBook2009}. From \eqref{Eq_3}, the densities of the PPPs $\Psi _{{\rm{LOS}}}$, $\Psi _{{\rm{NLOS}}}$ and $\Psi _{{\rm{OUT}}}$ are equal to $\lambda _{{\rm{LOS}}} \left( r \right) = \lambda p_{{\rm{LOS}}} \left( r \right)$, $\lambda _{{\rm{NLOS}}} \left( r \right) = \lambda p_{{\rm{NLOS}}} \left( r \right)$ and $\lambda _{{\rm{OUT}}} \left( r \right) = \lambda p_{{\rm{OUT}}} \left( r \right)$, respectively.
\subsection{Path-Loss Modeling} \label{PathLossModeling}
Based on the channel measurements reported in \cite{Rappaport__mmWaveJSAC}, the path-loss model for LOS and NLOS links is assumed to be:
\setcounter{equation}{3}
\begin{equation}
\label{Eq_4}
l_{{\rm{LOS}}} \left( r \right) = \left( {\kappa _{{\rm{LOS}}} r} \right)^{\beta _{{\rm{LOS}}} }, \; l_{{\rm{NLOS}}} \left( r \right) = \left( {\kappa _{{\rm{NLOS}}} r} \right)^{\beta _{{\rm{NLOS}}} }
\end{equation}
\noindent where $r$ denotes the generic BS-to-MT distance, $\kappa _{{\rm{LOS}}}$ and $\kappa _{{\rm{NLOS}}}$ can be interpreted as the path-loss of LOS and NLOS links at a distance of 1 meter, respectively, $\beta _{{\rm{LOS}}}$ and $\beta _{{\rm{NLOS}}}$ denote the power path-loss exponents of LOS and NLOS links, respectively.

The path-loss model in \eqref{Eq_4} is general enough for modeling several practical propagation conditions. For example, it can be linked to the widespread used $\left( {\alpha ,\beta } \right)$ model \cite{Rappaport__mmWaveAccess}, \cite{Rappaport__mmWaveJSAC}, by setting $\kappa _{{\rm{LOS}}}  = 10^{{{\alpha _{{\rm{LOS}}} } \mathord{\left/ {\vphantom {{\alpha _{{\rm{LOS}}} } {\left( {10\beta _{{\rm{LOS}}} } \right)}}} \right. \kern-\nulldelimiterspace} {\left( {10\beta _{{\rm{LOS}}} } \right)}}}$ and $\kappa _{{\rm{NLOS}}}  = 10^{{{\alpha _{{\rm{NLOS}}} } \mathord{\left/ {\vphantom {{\alpha _{{\rm{NLOS}}} } {\left( {10\beta _{{\rm{NLOS}}} } \right)}}} \right. \kern-\nulldelimiterspace} {\left( {10\beta _{{\rm{NLOS}}} } \right)}}}$, where $\alpha _{{\rm{LOS}}}$ and $\alpha _{{\rm{NLOS}}}$ are defined in \cite[Table I]{Rappaport__mmWaveJSAC}. As mentioned in Section \ref{LinkStateModeling}, the path-loss of the links that are in an outage state is assumed to be infinite, \textit{i.e.}, $l_{{\rm{OUT}}} \left( r \right) = \infty$.
\begin{figure*}[!t]
\setcounter{equation}{9}
\begin{equation} %\footnotesize
\label{Eq_10}
\begin{split}
 \Upsilon _0 \left( x;s \right) & = {\mathcal{K}}_1 \left( {1 - e^{ - Q_s x^{{1 \mathord{\left/
 {\vphantom {1 {\beta _s }}} \right.
 \kern-\nulldelimiterspace} {\beta _s }}} }  - Q_s x^{{1 \mathord{\left/
 {\vphantom {1 {\beta _s }}} \right.
 \kern-\nulldelimiterspace} {\beta _s }}} e^{ - Q_s x^{{1 \mathord{\left/
 {\vphantom {1 {\beta _s }}} \right.
 \kern-\nulldelimiterspace} {\beta _s }}} } } \right){\mathcal{\bar H}}\left( {x - Z_s } \right)
   + {\mathcal{K}}_1 \left( {1 - e^{ - R }  - R e^{ - R } } \right){\mathcal{H}}\left( {x - Z_s } \right) \\
   & + {\mathcal{K}}_2 \left( {e^{ - W }  + W e^{ - W }  - e^{ - V_s x^{{1 \mathord{\left/
 {\vphantom {1 {\beta _s }}} \right.
 \kern-\nulldelimiterspace} {\beta _s }}} }  - V_s x^{{1 \mathord{\left/
 {\vphantom {1 {\beta _s }}} \right.
 \kern-\nulldelimiterspace} {\beta _s }}} e^{ - V_s x^{{1 \mathord{\left/
 {\vphantom {1 {\beta _s }}} \right.
 \kern-\nulldelimiterspace} {\beta _s }}} } } \right){\mathcal{H}}\left( {x - Z_s } \right) \\
 \Upsilon _1 \left( {x;s} \right) & = \pi \lambda \kappa _s^{ - 2} x^{{2 \mathord{\left/
 {\vphantom {2 {\beta _s }}} \right.
 \kern-\nulldelimiterspace} {\beta _s }}} {\mathcal{\bar H}}\left( {x - Z_s } \right) + \pi \lambda \left( {\delta _{{\rm{OUT}}}^{ - 1} \ln \left( {\gamma _{{\rm{OUT}}} } \right)} \right)^2 {\mathcal{H}}\left( {x - Z_s } \right) \\
  & + 2\pi \lambda \delta _{{\rm{OUT}}}^{ - 2} \gamma _{{\rm{OUT}}} \left( {\gamma _{{\rm{OUT}}}^{ - 1}  + \gamma _{{\rm{OUT}}}^{ - 1} \ln \left( {\gamma _{{\rm{OUT}}} } \right) - e^{ - T_s x^{{1 \mathord{\left/
 {\vphantom {1 {\beta _s }}} \right.
 \kern-\nulldelimiterspace} {\beta _s }}} }  - T_s x^{{1 \mathord{\left/
 {\vphantom {1 {\beta _s }}} \right.
 \kern-\nulldelimiterspace} {\beta _s }}} e^{ - T_s x^{{1 \mathord{\left/
 {\vphantom {1 {\beta _s }}} \right.
 \kern-\nulldelimiterspace} {\beta _s }}} } } \right){\mathcal{H}}\left( {x - Z_s } \right) \\
 \end{split}
\end{equation}
\normalsize \hrulefill \vspace*{-5pt}
\end{figure*}
\subsection{Fading Modeling} \label{FadingModeling}
In addition to the distance-dependent path-loss model of Section \ref{PathLossModeling}, each link is assumed to be subject to a complex randomly distributed channel gain, which, for a generic BS-to-MT link, is denoted by $h$. According to \cite{Rappaport__mmWaveJSAC}, the power gain $\left| h \right|^2$ is assumed to follow a Log-Normal distribution with mean (in dB) equal to $\mu^{({\rm{dB}})}$ and standard deviation (in dB) equal to $\sigma^{({\rm{dB}})}$. Thus, $\left| h \right|^2$ takes into account large-scale shadowing. In general, $\mu^{({\rm{dB}})}$ and $\sigma^{({\rm{dB}})}$ for LOS and NLOS links are different \cite{Rappaport__mmWaveJSAC}. In what follows, they are denoted by $\mu _s^{\left( {{\rm{dB}}} \right)}$ and $\sigma _s^{\left( {{\rm{dB}}} \right)}$, where $s = \left\{ {{\rm{LOS}},{\rm{NLOS}}} \right\}$ denotes the link state.

As mentioned in Section \ref{LinkStateModeling}, for mathematical tractability, the shadowing correlations between links are ignored. Thus, the fading power gains of LOS and NLOS links are assumed to be independent and identically distributed. As recently remarked and verified with the aid of simulations in \cite{Heath__mmWave}, this assumption usually causes a minor loss of accuracy in the evaluation of the statistics of the Signal-to-Interference-plus-Noise-Ratio (SINR). Due to space limitations and for ease of description, fast-fading is neglected in the present paper, but it may be readily incorporated.
\subsection{Cell Association Criterion} \label{CellAssociation}
The MT is assumed to be served by the BS providing the smallest path-loss to it. Let $L_{{\rm{LOS}}}^{\left( 0 \right)}$, $L_{{\rm{NLOS}}}^{\left( 0 \right)}$ and $L_{{\rm{OUT}}}^{\left( 0 \right)}$ be the smallest path-loss of LOS, NLOS and OUT links, respectively. They can be formulated as:
\setcounter{equation}{4}
\begin{equation}
\label{Eq_5}
\begin{split}
& L_s^{\left( 0 \right)}  = \begin{cases}
 \mathop {\min }\limits_{n \in \Psi _s } \left\{ {l_s \left( {r^{\left( n \right)} } \right)} \right\}\quad & {\rm{if}}\quad \Psi _s  \ne \emptyset  \\
  + \infty \quad & {\rm{if}}\quad \Psi _s  = \emptyset  \\
 \end{cases} \\
 & L_{{\rm{OUT}}}^{\left( 0 \right)}  = \mathop {\min }\limits_{n \in \Psi _{{\rm{OUT}}} } \left\{ {l_{{\rm{OUT}}} \left( {r^{\left( n \right)} } \right)} \right\} =  + \infty
\end{split}
\end{equation}
\noindent where $s = \left\{ {{\mathop{\rm LOS}\nolimits} ,{\rm{NLOS}}} \right\}$, $r^{(n)}$ denotes the distance of a generic BS to the MT, and $\emptyset$ denotes an empty set. Hence, the path-loss of the serving BS, ${\rm{BS}}^{\left( 0 \right)}$, can be formulated as $L^{\left( 0 \right)}  = \min \left\{ {L_{{\rm{LOS}}}^{\left( 0 \right)} ,L_{{\rm{NLOS}}}^{\left( 0 \right)} ,L_{{\rm{OUT}}}^{\left( 0 \right)} } \right\}$.
\begin{figure*}[!t]
\setcounter{equation}{13}
\begin{equation}
\label{Eq_12}
\begin{split}
 \Upsilon _0^{\left( 1 \right)} \left( {x;s} \right) & = {\mathcal{K}}_2 \left( {e^{ - W}  + We^{ - W}  - e^{ - V_s x^{{1 \mathord{\left/
 {\vphantom {1 {\beta _s }}} \right.
 \kern-\nulldelimiterspace} {\beta _s }}} }  - V_s x^{{1 \mathord{\left/
 {\vphantom {1 {\beta _s }}} \right.
 \kern-\nulldelimiterspace} {\beta _s }}} e^{ - V_s x^{{1 \mathord{\left/
 {\vphantom {1 {\beta _s }}} \right.
 \kern-\nulldelimiterspace} {\beta _s }}} } } \right)\delta \left( {x - Z_s } \right) \\
 & - {\mathcal{K}}_1 \left( {1 - e^{ - Q_s x^{{1 \mathord{\left/
 {\vphantom {1 {\beta _s }}} \right.
 \kern-\nulldelimiterspace} {\beta _s }}} }  - Q_s x^{{1 \mathord{\left/
 {\vphantom {1 {\beta _s }}} \right.
 \kern-\nulldelimiterspace} {\beta _s }}} e^{ - Q_s x^{{1 \mathord{\left/
 {\vphantom {1 {\beta _s }}} \right.
 \kern-\nulldelimiterspace} {\beta _s }}} } } \right)\delta \left( {x - Z_s } \right)+ {\mathcal{K}}_1 \left( {1 - e^{ - R}  - Re^{ - R} } \right)\delta \left( {x - Z_s } \right) \\
 & + {\mathcal{K}}_2 \left( {{{V_s^2 } \mathord{\left/
 {\vphantom {{V_s^2 } {\beta _s }}} \right.
 \kern-\nulldelimiterspace} {\beta _s }}} \right)x^{{2 \mathord{\left/
 {\vphantom {2 {\beta _s }}} \right.
 \kern-\nulldelimiterspace} {\beta _s }} - 1} e^{ - V_s x^{{1 \mathord{\left/
 {\vphantom {1 {\beta _s }}} \right.
 \kern-\nulldelimiterspace} {\beta _s }}} } {\mathcal{H}}\left( {x - Z_s } \right) + {\mathcal{K}}_1 \left( {{{Q_s^2 } \mathord{\left/
 {\vphantom {{Q_s^2 } {\beta _s }}} \right.
 \kern-\nulldelimiterspace} {\beta _s }}} \right)x^{{2 \mathord{\left/
 {\vphantom {2 {\beta _s }}} \right.
 \kern-\nulldelimiterspace} {\beta _s }} - 1} e^{ - Q_s x^{{1 \mathord{\left/
 {\vphantom {1 {\beta _s }}} \right.
 \kern-\nulldelimiterspace} {\beta _s }}} } {\mathcal{\bar H}}\left( {x - Z_s } \right) \\
\Upsilon _1^{\left( 1 \right)} \left( {x;s} \right) & =  - \pi \lambda \kappa _s^{ - 2} x^{{2 \mathord{\left/
 {\vphantom {2 {\beta _s }}} \right.
 \kern-\nulldelimiterspace} {\beta _s }}} \delta \left( {x - Z_s } \right) + \pi \lambda \left( {\delta _{{\rm{OUT}}}^{ - 1} \ln \left( {\gamma _{{\rm{OUT}}} } \right)} \right)^2 \delta \left( {x - Z_s } \right) \\
 & + 2\pi \lambda \delta _{{\rm{OUT}}}^{ - 2} \gamma _{{\rm{OUT}}} \left( {\gamma _{{\rm{OUT}}}^{ - 1}  + \gamma _{{\rm{OUT}}}^{ - 1} \ln \left( {\gamma _{{\rm{OUT}}} } \right) - e^{ - T_s x^{{1 \mathord{\left/
 {\vphantom {1 {\beta _s }}} \right.
 \kern-\nulldelimiterspace} {\beta _s }}} }  - T_s x^{{1 \mathord{\left/
 {\vphantom {1 {\beta _s }}} \right.
 \kern-\nulldelimiterspace} {\beta _s }}} e^{ - T_s x^{{1 \mathord{\left/
 {\vphantom {1 {\beta _s }}} \right.
 \kern-\nulldelimiterspace} {\beta _s }}} } } \right)\delta \left( {x - Z_s } \right) \\
 & + 2\pi \lambda \kappa _s^{ - 2} \beta _s^{ - 1} x^{{2 \mathord{\left/
 {\vphantom {2 {\beta _s }}} \right.
 \kern-\nulldelimiterspace} {\beta _s }} - 1} {\mathcal{\bar H}}\left( {x - Z_s } \right) + 2\pi \lambda \delta _{{\rm{OUT}}}^{ - 2} \gamma _{{\rm{OUT}}} T_s^2 \beta _s^{ - 1} x^{{2 \mathord{\left/
 {\vphantom {2 {\beta _s }}} \right.
 \kern-\nulldelimiterspace} {\beta _s }} - 1} e^{ - T_s x^{{1 \mathord{\left/
 {\vphantom {1 {\beta _s }}} \right.
 \kern-\nulldelimiterspace} {\beta _s }}} } {\mathcal{H}}\left( {x - Z_s } \right) \\
\end{split}
\end{equation}
\normalsize \hrulefill \vspace*{-5pt}
\end{figure*}
\section{Mathematical Modeling of Coverage Probability and Average Rate} \label{CoverageRate}
Based on the system model of Section \ref{SystemModel}, the SINR of the cellular network under analysis can be formulated as follows:
\setcounter{equation}{5}
\begin{equation}
\label{Eq_6}
\begin{split}
 {\rm{SINR}} & = \frac{{{{PG^{\left( 0 \right)} \left| {h^{\left( 0 \right)} } \right|^2 } \mathord{\left/
 {\vphantom {{PG^{\left( 0 \right)} \left| {h^{\left( 0 \right)} } \right|^2 } {L^{\left( 0 \right)} }}} \right.
 \kern-\nulldelimiterspace} {L^{\left( 0 \right)} }}}}{{\sigma _N^2  + I_{{\rm{agg}}} }} \\ & \mathop  = \limits^{\left( a \right)} \frac{{{{PG^{\left( 0 \right)} \left| {h_{{\rm{LOS}}}^{\left( {\rm{0}} \right)} } \right|^2 } \mathord{\left/
 {\vphantom {{PG^{\left( 0 \right)} \left| {h_{{\rm{LOS}}}^{\left( {\rm{0}} \right)} } \right|^2 } {L^{\left( 0 \right)} }}} \right.
 \kern-\nulldelimiterspace} {L^{\left( 0 \right)} }}}}{{\sigma _N^2  + I_{{\rm{agg}}} }}\delta \left\{ {L^{\left( 0 \right)}  - L_{{\rm{LOS}}}^{\left( 0 \right)} } \right\} \\
 & + \frac{{{{PG^{\left( 0 \right)} \left| {h_{{\rm{NLOS}}}^{\left( {\rm{0}} \right)} } \right|^2 } \mathord{\left/
 {\vphantom {{PG^{\left( 0 \right)} \left| {h_{{\rm{NLOS}}}^{\left( {\rm{0}} \right)} } \right|^2 } {L^{\left( 0 \right)} }}} \right.
 \kern-\nulldelimiterspace} {L^{\left( 0 \right)} }}}}{{\sigma _N^2  + I_{{\rm{agg}}} }}\delta \left\{ {L^{\left( 0 \right)}  - L_{{\rm{NLOS}}}^{\left( 0 \right)} } \right\}
 \end{split}
\end{equation}
\noindent where $P$ is the transmit power, $\left| {h^{\left( 0 \right)} } \right|^2$ is the power gain of the serving BS, ${\rm{BS}}^{\left( 0 \right)}$, $\sigma _N^2$ is the noise power and $I_{{\rm{agg}}}$ is the aggregate other-cell interference, \textit{i.e.}, the total interference generated by the BSs in $\Psi ^{\left(\backslash 0 \right)}$. In particular, $\sigma _N^2$ is defined as $\sigma _N^2  = 10^{{{\sigma _N^2 \left( {{\rm{dBm}}} \right)} \mathord{\left/ {\vphantom {{\sigma _N^2 \left( {{\rm{dBm}}} \right)} {10}}} \right. \kern-\nulldelimiterspace} {10}}}$, where $\sigma _N^2 \left( {{\rm{dBm}}} \right) =  - 174 + 10\log _{10} \left( {{\rm{BW}}} \right) + \mathcal{F}_{{\rm{dB}}}$, ${{\rm{BW}}}$ is the transmission bandwidth and $\mathcal{F}_{{\rm{dB}}}$ is the noise figure in dB. By using a notation similar to the desired link, the aggregate other-cell interference is defined as $I_{{\rm{agg}}}  = \sum\nolimits_{i \in \Psi ^{\left( {\backslash 0} \right)} } {\left( {{{P G^{\left( i \right)} \left| {h^{\left( i \right)} } \right|^2 } \mathord{\left/ {\vphantom {{P G^{\left( i \right)} \left| {h^{\left( i \right)} } \right|^2 } {L^{\left( i \right)} }}} \right. \kern-\nulldelimiterspace} {L^{\left( i \right)} }}} \right)}$. The equality in (a) takes into account that the distribution of LOS and NLOS links is assumed to be different in the present paper.

From \eqref{Eq_6}, coverage probability (${\rm{P}}_{{\mathop{\rm cov}} }$) and average rate ($\rm{R}$) can be formulated as follows \cite{MDR_COMMLPeng}:
\setcounter{equation}{6}
\begin{equation}
\label{Eq_7}
{\rm{P}}^{({\mathop{\rm cov}} )} \left( {\rm{T}} \right) = \Pr \left\{ {{\rm{SINR}} \ge {\rm{T}}} \right\}
\end{equation}
\setcounter{equation}{7}
\begin{equation}
\label{Eq_8}
\begin{split}
{\rm{R}} &= {\mathbb{E}}_{{\rm{SINR}}} \left\{ {{\rm{BW}}\log _2 \left( {1 + {\rm{SINR}}} \right)} \right\} \\ & = \frac{{{\rm{BW}}}}{{\ln \left( 2 \right)}}\int\nolimits_0^{ + \infty } {\frac{{{\rm{P}}_{{\mathop{\rm cov}} } \left( t \right)}}{{t + 1}}dt}
\end{split}
\end{equation}
\noindent where ${\rm{T}}$ is a reliability threshold and ${\mathbb{E}}\left\{  \cdot  \right\}$ denotes the expectation operator.

In what follows, a new mathematical expression for the coverage probability, ${\rm{P}}^{({\mathop{\rm cov}})}(\cdot)$, is provided. The average rate, $\rm{R}$, can be obtained directly from \eqref{Eq_8}. The analytical formulation is based on the noise-limited approximation of mmWave cellular communications, \textit{i.e.}, ${\rm{SINR}} \approx {\rm{SNR}} = P G^{\left( 0 \right)} \left| {h^{\left( 0 \right)} } \right|^2 \left( {\sigma _N^2 L^{\left( 0 \right)} } \right)^{-1}$, which has been observed in recent studies, both with the aid of numerical simulations and field measurements \cite{Rappaport__mmWaveJSAC}, \cite{Singh__mmWave}. In Section \ref{Results}, the validity and the accuracy of the noise-limited approximation are substantiated with the aid of Monte Carlo simulations, which account for the other-cell interference as well.

Before presenting the main result summarized in \textit{Proposition \ref{Pcov_Prop}}, we introduce two lemmas that are useful for its proof.
\begin{lemma} \label{Intensity_Lemma}
Let $L  = \left\{ {L _{{\rm{LOS}}} ,L _{{\rm{NLOS}}} ,L _{{\rm{OUT}}} } \right\}$, where $L_{\bar s}  = \left\{ {l_{\bar s} \left( {r^{\left( n \right)} } \right),n \in \Psi _{\bar s} } \right\}$ for $\bar s \in \left\{ {{\rm{LOS}},{\rm{NLOS}},{\rm{OUT}}} \right\}$ are transformations of the path-loss of LOS, NLOS and OUT BSs, respectively. Let the link state and the path-loss models in \eqref{Eq_3} and \eqref{Eq_4}, respectively. $L$ is a PPP with intensity as follows:
\setcounter{equation}{8}
\begin{equation}
\label{Eq_9}
\Lambda \left( {\left[ {0,x} \right)} \right) = \Lambda _{{\rm{LOS}}} \left( {\left[ {0,x} \right)} \right) + \Lambda _{{\rm{NLOS}}} \left( {\left[ {0,x} \right)} \right)
\end{equation}
\noindent where $\Lambda _{{\rm{LOS}}} \left( {\left[ {0,x} \right)} \right) = \Upsilon _0 \left( {x;s = {\rm{LOS}}} \right)$, $\Lambda _{{\rm{NLOS}}} \left( {\left[ {0,x} \right)} \right) = \Upsilon _1 \left( {x;s = {\rm{NLOS}}} \right) - \Upsilon _0 \left( {x;s = {\rm{NLOS}}} \right)$, $\Upsilon _0 \left( \cdot;\cdot \right)$ and $\Upsilon _1 \left( \cdot;\cdot \right)$ are defined in \eqref{Eq_10}, ${\mathcal{H}}\left(  \cdot  \right)$ is the Heaviside function, ${\mathcal{\bar H}}\left( x \right) = 1 - {\mathcal{H}}\left( x \right)$, ${\mathcal{K}}_1  = 2\pi \lambda \gamma _{{\rm{LOS}}} \delta _{{\rm{LOS}}}^{ - 2}$, ${\mathcal{K}}_2  = 2\pi \lambda \gamma _{{\rm{LOS}}} \gamma _{{\rm{OUT}}} \left( {\delta _{{\rm{LOS}}}  + \delta _{{\rm{OUT}}} } \right)^{ - 2}$, $R  = \delta _{{\rm{LOS}}} \delta _{{\rm{OUT}}}^{ - 1} \ln \left( {\gamma _{{\rm{OUT}}} } \right)$, $W  = \left( {\delta _{{\rm{LOS}}}  + \delta _{{\rm{OUT}}} } \right)\delta _{{\rm{OUT}}}^{ - 1} \ln \left( {\gamma _{{\rm{OUT}}} } \right)$, $Q_s  = \delta _{{\rm{LOS}}} \kappa _s^{ - 1}$, $T_s  = \delta _{{\rm{OUT}}} \kappa _s^{ - 1}$, $V_s  = \left( {\delta _{{\rm{LOS}}}  + \delta _{{\rm{OUT}}} } \right)\kappa _s^{ - 1}$, $Z_s  = \left( {\kappa _s \delta _{{\rm{OUT}}}^{ - 1} \ln \left( {\gamma _{{\rm{OUT}}} } \right)} \right)^{\beta _s }$ for $s = \left\{ {{\rm{LOS}},{\rm{NLOS}}} \right\}$.

\emph{Proof}: See the Appendix. \hfill $\Box$
\end{lemma}
\begin{corollary} \label{CDF_Corollary}
Let $\delta _{{\rm{OUT}}}  = 0$, $\gamma _{{\rm{OUT}}}  = 1$, \textit{i.e.}, $p_{{\rm{OUT}}} \left( r \right) = 0$ in \eqref{Eq_3}. $\Lambda  \left(  \cdot  \right)$ in \eqref{Eq_9} holds with $\Upsilon _0 \left( x;s \right) = {\mathcal{K}}_1 \left( {1 - e^{ - Q_s x^{{1 \mathord{\left/ {\vphantom {1 {\beta _s }}} \right. \kern-\nulldelimiterspace} {\beta _s }}} }  - Q_s x^{{1 \mathord{\left/ {\vphantom {1 {\beta _s }}} \right. \kern-\nulldelimiterspace} {\beta _s }}} e^{ - Q_s x^{{1 \mathord{\left/ {\vphantom {1 {\beta _s }}} \right. \kern-\nulldelimiterspace} {\beta _s }}} } } \right)$, $\Upsilon _1 \left( {x;s} \right) = \pi \lambda \kappa _s^{ - 2} x^{{2 \mathord{\left/ {\vphantom {2 {\beta _s }}} \right. \kern-\nulldelimiterspace} {\beta _s }}}$, $s = \left\{ {{\rm{LOS}},{\rm{NLOS}}} \right\}$.

\emph{Proof}: It follows directly from \eqref{Eq_10}, since $Z_s \to +\infty$ for $s = \left\{ {{\rm{LOS}},{\rm{NLOS}}} \right\}$. \hfill $\Box$
\end{corollary}
\begin{lemma} \label{CDF_Lemma}
Let $L ^{\left( 0 \right)} = \min \left\{ {L_{{\rm{LOS}}}^{\left( 0 \right)} ,L_{{\rm{NLOS}}}^{\left( 0 \right)} ,L_{{\rm{OUT}}}^{\left( 0 \right)} } \right\} = \min \left\{ L  \right\}$ be the smallest element of the PPP $L$ introduced in \textit{Lemma \ref{Intensity_Lemma}}. Its Cumulative Distribution Function (CDF), \textit{i.e.}, $F_{L^{\left( 0 \right)} } \left( x \right) = \Pr \left\{ {L^{\left( 0 \right)}  \le x} \right\}$, can be formulated as follows:
\setcounter{equation}{10}
\begin{equation}
\label{Eq_11}
F_{L^{\left( 0 \right)} } \left( x \right) = 1 - \exp \left( { - \Lambda \left( {\left[ {0,x} \right)} \right)} \right)
\end{equation}
\noindent where $\Lambda \left( \cdot \right)$ is defined in \eqref{Eq_9}.

\emph{Proof}: It follows by applying the void probability theorem of PPPs \cite[Corollary 6]{Blaszczyszyn_Infocom2013}. \hfill $\Box$
\end{lemma}
\subsection{Coverage Probability}
\begin{proposition} \label{Pcov_Prop}
Let the SINR in \eqref{Eq_6} and the coverage probability in \eqref{Eq_7}. If $\sigma _N^2  \gg I_{{\rm{agg}}}$, \textit{i.e.}, ${\rm{SINR}} \approx {\rm{SNR}} = P G^{\left( 0 \right)} \left| {h^{\left( 0 \right)} } \right|^2 \left( {\sigma _N^2 L^{\left( 0 \right)} } \right)^{-1}$, then the following holds:
\setcounter{equation}{11}
\begin{equation}
\label{Eq_13}
{\rm{P}}^{\left( {{\mathop{\rm cov}} } \right)} \left( {\rm{T}} \right) = {\rm{P}}_{{\rm{LOS}}}^{\left( {{\mathop{\rm cov}} } \right)} \left( {\rm{T}} \right) + {\rm{P}}_{{\rm{NLOS}}}^{\left( {{\mathop{\rm cov}} } \right)} \left( {\rm{T}} \right)
\end{equation}
\setcounter{equation}{12}
\begin{equation}
\label{Eq_14}
\hspace{-0.15cm} {\rm{P}}_s^{\left( {{\mathop{\rm cov}} } \right)} \left( {\rm{T}} \right) = \frac{1}{2} \int\nolimits_0^{ + \infty } {{\rm{erfc}}\left( {\frac{{{\rm{ln}}\left( {{{{\rm{T}}x} \mathord{\left/
 {\vphantom {{{\rm{T}}x} {\gamma ^{\left( 0 \right)} }}} \right.
 \kern-\nulldelimiterspace} {\gamma ^{\left( 0 \right)} }}} \right) - \mu _s }}{{\sqrt 2 \sigma _s }}}  \right) \bar \Lambda_s \left( {\left[ {0,x} \right)} \right) dx}
\end{equation}
\noindent where $s = \left\{ {{\rm{LOS}},{\rm{NLOS}}} \right\}$, ${\rm{erfc}}\left(  \cdot  \right)$ is the complementary error function, $\bar \Lambda_s \left( {\left[ {0,x} \right)} \right)  = \Lambda _{s }^{\left( 1 \right)} \left( {\left[ {0,x} \right)} \right)\exp \left( { - \Lambda \left( {\left[ {0,x} \right)} \right)} \right)$, $\Lambda \left( \cdot \right)$ is defined in \eqref{Eq_9}, $\Lambda _{{{\rm{LOS}}} }^{\left( 1 \right)} \left( {\left[ {0,x} \right)} \right) = \Upsilon _0^{\left( 1 \right)} \left( {x;{\rm{LOS}}} \right)$, where $\Upsilon _0^{\left( 1 \right)} \left( { \cdot ; \cdot } \right)$ is the first derivative of $\Upsilon _0 \left( { \cdot ; \cdot } \right)$ in \eqref{Eq_10}, $\Lambda _{{{\rm{NLOS}}} }^{\left( 1 \right)} \left( {\left[ {0,x} \right)} \right) = \Upsilon _1^{\left( 1 \right)} \left( {x;{\rm{NLOS}}} \right) - \Upsilon _0^{\left( 1 \right)} \left( {x;{\rm{NLOS}}} \right)$, where $\Upsilon _1^{\left( 1 \right)} \left( { \cdot ; \cdot } \right)$ is the first derivative of $\Upsilon _1 \left( { \cdot ; \cdot } \right)$ in \eqref{Eq_10}, $\gamma ^{\left( 0 \right)}  = {{{{P}}G^{\left( 0 \right)} } \mathord{\left/ {\vphantom {{{\mathsf{P}}G^{\left( 0 \right)} } {\sigma _N^2 }}} \right. \kern-\nulldelimiterspace} {\sigma _N^2 }}$, $\mu _s  = \mu _s^{\left( {{\rm{dB}}} \right)} {{\ln \left( {10} \right)} \mathord{\left/ {\vphantom {{\ln \left( {10} \right)} {10}}} \right. \kern-\nulldelimiterspace} {10}}$ and $\sigma _s  = \sigma _s^{\left( {{\rm{dB}}} \right)} {{\ln \left( {10} \right)} \mathord{\left/ {\vphantom {{\ln \left( {10} \right)} {10}}} \right. \kern-\nulldelimiterspace} {10}}$. $\Upsilon _0^{\left( 1 \right)} \left( { \cdot ; \cdot } \right)$ and $\Upsilon _1^{\left( 1 \right)} \left( { \cdot ; \cdot } \right)$ are available in \eqref{Eq_12}.

\emph{Proof}: See the Appendix. \hfill $\Box$
\end{proposition}
\section{Numerical and Simulation Results} \label{Results}
In this section, we illustrate some numerical examples for validating the accuracy of the proposed mathematical frameworks and for comparing mmWave and $\mu$Wave cellular networks. The results are compared against Monte Carlo simulations, where some modeling assumptions considered for developing the mathematical frameworks are not enforced in the system simulator. Notably, coverage probability and average rate in Section \ref{CoverageRate} are computed under the assumption of noise-limited cellular networks. This assumption is \textit{not} retained in the system simulator, in order to show to which extent the noise-limited assumption holds for mmWave cellular communications. Monte Carlo simulation results are obtained by using the system simulator described in \cite{MDR_TCOMrate}-\cite{MDR_COMMLPeng}, to which the reader is referred for further information.

Unless otherwise stated, the following setup is considered for obtaining the numerical examples, which agrees with previous studies in this field \cite{Rappaport__mmWaveJSAC}, \cite{Heath__mmWave}, \cite{Singh__mmWave}. In particular, the channel and blockage models are taken from \cite{Rappaport__mmWaveJSAC}. In addition:
\begin{itemize}
  \item Two mmWave cellular networks are studied, which operate at a carrier frequency, $F_c$, equal to $F_c=28$ GHz and $F_c=73$ GHz. The transmission bandwidth is ${\rm{BW}} = 2$ GHz. The noise figure is ${\mathcal{N}}_{\rm{dB}}  = 10$. The transmit power is $P=30$ dBm.
  \item The path-loss model is as follows \cite[Table I]{Rappaport__mmWaveJSAC}: $\alpha _{{\rm{LOS}}}  = 61.4$ dB, $\beta _{{\rm{LOS}}}  = 2$ and $\alpha _{{\rm{NLOS}}}  = 72$ dB, $\beta _{{\rm{NLOS}}}  = 2.92$ if $F_c=28$ GHz and $\alpha _{{\rm{LOS}}}  = 69.8$ dB, $\beta _{{\rm{LOS}}}  = 2$ and $\alpha _{{\rm{NLOS}}}  = 82.7$ dB, $\beta _{{\rm{NLOS}}}  = 2.69$ if $F_c=73$ GHz.
  \item The shadowing model is as follows \cite[Table I]{Rappaport__mmWaveJSAC}: $\sigma _{{\rm{LOS}}}^{(\rm{dB})}  = 5.8$, $\sigma _{{\rm{NLOS}}}^{(\rm{dB})}  = 8.7$ if $F_c=28$ GHz and $\sigma _{{\rm{LOS}}}^{(\rm{dB})}  = 5.8$, $\sigma _{{\rm{NLOS}}}^{(\rm{dB})}  = 8.7$ if $F_c=73$ GHz. On the other hand, $\mu^{(\rm{dB})}$ is assumed to be equal to zero for both LOS and NLOS scenarios.
  \item The blockage model is as follows \cite[Table I]{Rappaport__mmWaveJSAC}: $\delta _{{\rm{LOS}}}  = {1 \mathord{\left/ {\vphantom {1 {67.1}}} \right. \kern-\nulldelimiterspace} {67.1}}$, $\gamma _{{\rm{LOS}}}  = 1$ and $\delta _{{\rm{OUT}}}  = 5.2$, $\gamma _{{\rm{OUT}}}  = \exp \left( {{1 \mathord{\left/ {\vphantom {1 {30}}} \right. \kern-\nulldelimiterspace} {30}}} \right)$, for both $F_c=28$ GHz and $F_c=73$ GHz scenarios.
  \item The directional beamforming model is as follows \cite{Heath__mmWave}: $G_{{\rm{BS}}}^{\left( {\max } \right)}  = G_{{\rm{MT}}}^{\left( {\max } \right)}  = 20$ dB, $G_{{\rm{BS}}}^{\left( {\min } \right)}  = G_{{\rm{MT}}}^{\left( {\min } \right)}  = -10$ dB and $\omega _{{\rm{BS}}}  = \omega _{{\rm{MT}}}  = 30$ degrees.
  \item Similar to \cite{Heath__mmWave}, the density of BSs, $\lambda$, is represented as a function of the average cell radius, \textit{i.e.}, $R_c  = \sqrt {{1 \mathord{\left/{\vphantom {1 {\left( {\pi \lambda } \right)}}} \right. \kern-\nulldelimiterspace} {\left( {\pi \lambda } \right)}}}$.
  \item As for the $\mu$Wave cellular network, a setup similar to \cite{Rappaport__mmWaveJSAC} is considered. In particular, we set $F_c = 2.5$ GHz, ${\rm{BW}} = 40$ MHz, $G_{{\rm{MT}}}^{\left( {\max } \right)}  = G_{{\rm{MT}}}^{\left( {\min } \right)}  = 0$ dB and $\omega _{{\rm{MT}}}  = 360$ degrees. The channel model is chosen as in \cite[Eq. (11)]{Rappaport__mmWaveJSAC}, \textit{i.e.}, $l\left( r \right)^{\left( {{\rm{dB}}} \right)}  = 22.7 + 36.7\log _{10} \left( r \right) + 26\log _{10} \left( {2.5} \right)$. All the channels are assumed to be in a NLOS state, with a shadowing standard deviation equal to $\sigma _{{\rm{NLOS}}}  = 4$. No outage state is assumed, \textit{i.e.}, $p_{{\rm{OUT}}} \left( r \right) = 0$. The rest of the paraments is the same as for the mmWave cellular network setup.
\end{itemize}
\begin{figure}[!t]
\centering
\includegraphics [width=\columnwidth] {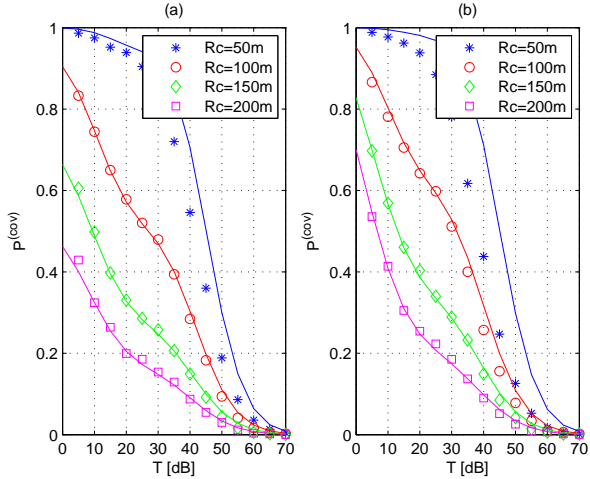}
\vspace{-0.75cm} \caption{Coverage probability of a mmWave cellular network operating at $F_c=28$ GHz. (a) $p_{{\rm{OUT}}}(\cdot)$ in \eqref{Eq_3} is used. (b) $p_{{\rm{OUT}}}(r) = 0$. Solid lines show the mathematical framework and markers show Monte Carlo simulations.}
\label{Fig_1}
\end{figure}
\begin{figure}[!t]
\centering
\includegraphics [width=\columnwidth] {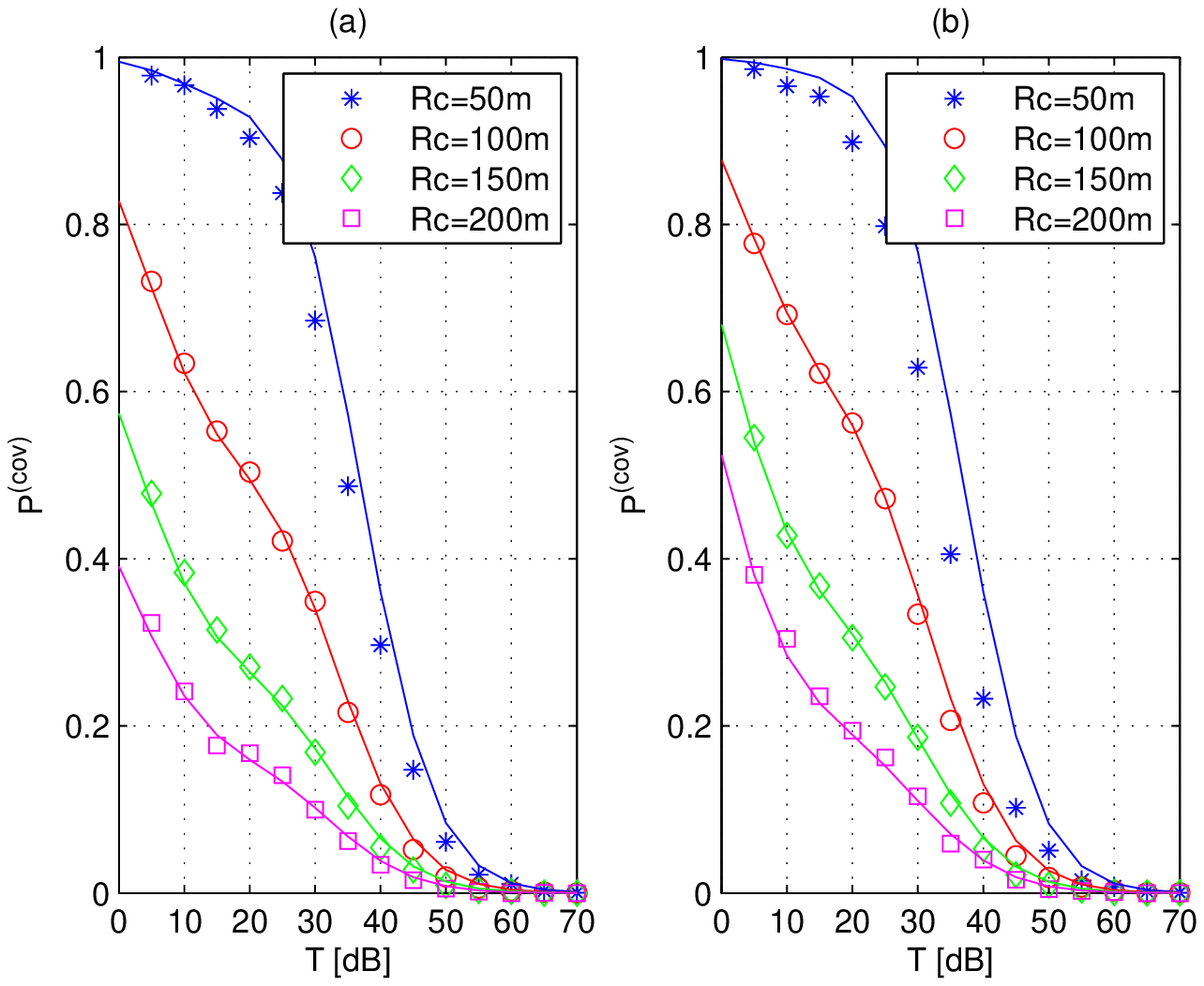}
\vspace{-0.75cm} \caption{Coverage probability of a mmWave cellular network operating at $F_c=73$ GHz. (a) $p_{{\rm{OUT}}}(\cdot)$ in \eqref{Eq_3} is used. (b) $p_{{\rm{OUT}}}(r) = 0$. Solid lines show the mathematical framework and markers show Monte Carlo simulations.} \label{Fig_2}
\end{figure}
\begin{figure}[!t]
\centering
\includegraphics [width=\columnwidth] {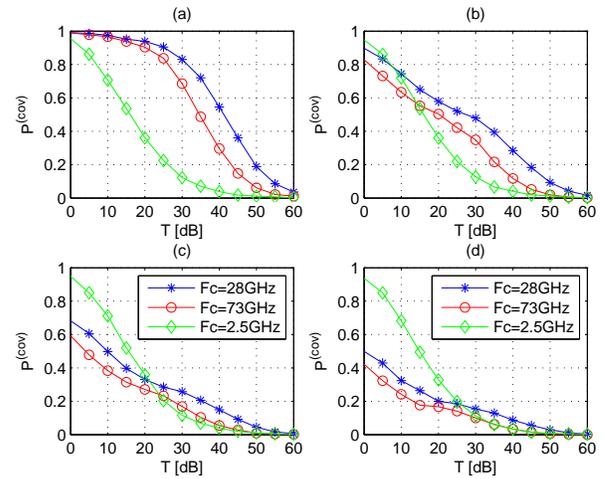}
\vspace{-0.75cm} \caption{Coverage probability of mmWave and $\mu$Wave cellular networks operating at $F_c=28$ GHz (mmWave), $F_c=73$ GHz (mmWave) and $F_c=2.5$ GHz ($\mu$Wave). For mmWave networks, $p_{{\rm{OUT}}}(\cdot)$ in \eqref{Eq_3} is used. For the $\mu$Wave cellular network, $p_{{\rm{OUT}}}(r) = 0$. (a) $R_c = 50$ m. (b) $R_c = 100$ m. (c) $R_c = 150$ m. (d) $R_c = 200$ m.} \label{Fig_3}
\end{figure}
\begin{figure}[!t]
\centering
\includegraphics [width=\columnwidth] {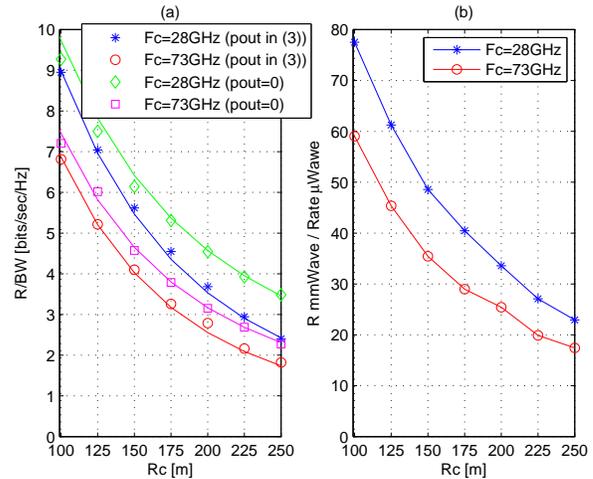}
\vspace{-0.75cm} \caption{Average rate of a mmWave cellular network operating at $F_c=28$ GHz and $F_c=73$ GHz. (a) The normalized rate ${\rm{R}}/{\rm{BW}}$ is shown. Solid lines show the mathematical framework and markers show Monte Carlo simulations. (b) Ratio of the average rates of two mmWave networks operating at $F_c=28$ GHz and $F_c=73$ GHz and of a $\mu$Wave network operating at $F_c=2.5$ GHz.} \label{Fig_4}
\end{figure}
Some selected numerical results are illustrated in Figs. \ref{Fig_1}-\ref{Fig_4}. From Figs. \ref{Fig_1} and \ref{Fig_2}, we observe that the proposed noise-limited approximation is quite accurate for practical densities of BSs. If $R_c \ge 100$ meters for the considered setup, in particular, we observe that mmWave cellular networks can be assumed to be noise-limited. If the density of BSs increases, on the other hand, this approximation may no longer hold. The performance gap compared to Monte Carlo simulations is, however, tolerable and this shows that, in any case, mmWave cellular networks are likely not to be interference-limited. This finding is in agreement with recent published papers that considered a simplified blockage model \cite{Singh__mmWave}. The figures also show that, in general, the presence of an outage state reduces the coverage probability. This is noticeable, however, only for small values of the reliability threshold $\rm{T}$.

In Fig. \ref{Fig_3}, mmWave and $\mu$Wave cellular networks are compared. This figure shows that mmWave communications have the potential of outperforming $\mu$Wave communications, provided that the network density is sufficiently high. Otherwise, $\mu$Wave communications are still to be preferred, especially for small values of the reliability threshold $\rm{T}$. As expected, mmWave transmission at $F_c=28$ GHz slightly outperforms its counterpart at $F_c=73$ GHz due to a smaller path-loss.

Finally, Fig. \ref{Fig_4} shows the average rate of mmWave and $\mu$Wave cellular networks. It confirms that the proposed framework is accurate enough for practical network densities. Also, it highlights that mmWave communications are capable of significantly enhancing the average rate. This is mainly due to the larger transmission bandwidth, which is 50 times larger, in the considered setup, for mmWave communications. The figure shows, however, that the gain can be larger than the ratio of the bandwidths, especially for medium/dense cellular network deployments.
\begin{figure*}[!t]
\setcounter{equation}{15}
\begin{equation}
\label{Eq_App2}
\begin{split}
 {\rm{P}}^{\left( {{\mathop{\rm cov}} } \right)} \left( {\rm{T}} \right) & = {\mathbb{E}}_{L_{{\rm{LOS}}}^{\left( 0 \right)} } \left\{ {\Pr \left\{ {\left. {\frac{{{{P}}G^{\left( 0 \right)} \left| {h_{{\rm{LOS}}}^{\left( 0 \right)} } \right|^2 }}{{\sigma _N^2 L_{{\rm{LOS}}}^{\left( 0 \right)} }} > {\rm{T}}} \right|L_{{\rm{LOS}}}^{\left( 0 \right)} } \right\}\Pr \left\{ {\left. {L_{{\rm{NLOS}}}^{\left( 0 \right)}  > L_{{\rm{LOS}}}^{\left( 0 \right)} } \right|L_{{\rm{LOS}}}^{\left( 0 \right)} } \right\}} \right\} \\
 & + {\mathbb{E}}_{L_{{\rm{NLOS}}}^{\left( 0 \right)} } \left\{ {\Pr \left\{ {\left. {\frac{{{{P}}G^{\left( 0 \right)} \left| {h_{{\rm{NLOS}}}^{\left( 0 \right)} } \right|^2 }}{{\sigma _N^2 L_{{\rm{NLOS}}}^{\left( 0 \right)} }} > {\rm{T}}} \right|L_{{\rm{NLOS}}}^{\left( 0 \right)} } \right\}\Pr \left\{ {\left. {L_{{\rm{LOS}}}^{\left( 0 \right)}  > L_{{\rm{NLOS}}}^{\left( 0 \right)} } \right|L_{{\rm{NLOS}}}^{\left( 0 \right)} } \right\}} \right\} \\
 \end{split}
\end{equation}
\normalsize \hrulefill \vspace*{-5pt}
\end{figure*}
\section{Conclusion} \label{Conclusion}
In the present paper, a new analytical framework for computing coverage probability and average rate of mmWave cellular networks has been proposed. Its novelty lies in taking into account realistic channel and blockage models for mmWave propagation, which are based on empirical data recently reported in the literature. The proposed methodology relies upon the noise-limited assumption for mmWave communications, which is shown to be sufficiently accurate for typical densities of BSs. The numerical examples have shown that sufficiently dense mmWave cellular networks have the inherent capability of outperforming $\mu$Wave cellular networks.

Current research activities are concerned with the extension of the proposed mathematical framework to ultra-dense mmWave cellular networks, where interference can no longer be neglected, as well as to compare different blockage models and cell association criteria. An extended version of the present paper is available in \cite{MDR_TWCmmWave}.
%
%
%
%\appendices
%
%
\section*{Appendix -- Proofs} \label{Appendix}
\subsection{Proof of Lemma \ref{Intensity_Lemma}}
The proof follows by using a methodology similar to \cite[Sec. II-A]{Blaszczyszyn_Infocom2013}. In particular, by invoking the displacement theorem of PPPs \cite[Th. 1.10]{BaccelliBook2009}, the process of the propagation losses $L = \left\{ {l\left( {r^{(n)} } \right)}, n \in \Psi \right\}$ can be interpreted as a transformation of $\Psi$, which is still a PPP on $\mathbb{R}^+$. From Section \ref{LinkStateModeling}, we know that $\Psi  = \Psi _{{\rm{LOS}}}  \cup \Psi _{{\rm{NLOS}}}  \cup \Psi _{{\rm{OUT}}}$. Since $\Psi _{{\rm{LOS}}}$, $\Psi _{{\rm{NLOS}}}$ and $\Psi _{{\rm{OUT}}}$ are independent, the density (or intensity), ${\Lambda \left( \cdot \right)}$, of $L$ is equal to the summation of the intensities of $\Psi _{{\rm{LOS}}}$, $\Psi _{{\rm{NLOS}}}$ and $\Psi _{{\rm{OUT}}}$, \textit{i.e.}, $\Lambda \left( {\left[ {0,x} \right)} \right) = \Lambda _{{\rm{LOS}}} \left( {\left[ {0,x} \right)} \right) + \Lambda _{{\rm{NLOS}}} \left( {\left[ {0,x} \right)} \right) + \Lambda _{{\rm{OUT}}} \left( {\left[ {0,x} \right)} \right)$. Since the path-loss of the links in outage is infinite, by definition $\Lambda _{{\rm{OUT}}} \left( {\left[ {0,x} \right)} \right) = 0$. On the other hand, $\Lambda _s \left( \cdot \right)$ for $s = \left\{ {{\rm{LOS}},{\rm{NLOS}}} \right\}$ can be computed by using mathematical steps similar to the proof of \cite[Lemma 1]{Blaszczyszyn_Infocom2013}. More specifically, we have:
\setcounter{equation}{14}
\begin{equation}
\label{Eq_App1}
\Lambda _s \left( {\left[ {0,x} \right)} \right) = 2\pi \lambda \int\nolimits_0^{ + \infty } {{\mathcal{H}}\left( {x - \left( {\kappa _s r} \right)^{\beta _s } } \right)p_s \left( r \right)rdr}
\end{equation}

Equation \eqref{Eq_9} follows by inserting $p_s \left(  \cdot  \right)$ of \eqref{Eq_3} in \eqref{Eq_App1} and by computing the integrals with the aid of the notable result $\int\nolimits_a^b {e^{ - cr} rdr}  = \left( {{1 \mathord{\left/ {\vphantom {1 {c^2 }}} \right.\kern-\nulldelimiterspace} {c^2 }}} \right)\left( {e^{ - ca}  + ae^{ - ca}  - e^{ - cb}  - be^{ - cb} } \right)$.
\subsection{Proof of Proposition \ref{Pcov_Prop}}
From \eqref{Eq_6} and \eqref{Eq_7}, the coverage probability can be formulated, by definition, as shown in \eqref{Eq_App2}. Let us denote the first and second addends in \eqref{Eq_App2} by ${\rm{P}}_s^{\left( {{\mathop{\rm cov}} } \right)} \left( \cdot \right)$, where $s = {\rm{LOS}}$ and $s = {\rm{NLOS}}$, respectively. Each addend can be computed by using the following results:
\setcounter{equation}{16}
\begin{equation}
\label{Eq_App3}
\hspace{-0.1cm}\begin{split}
 & \Pr \left\{ {\left. {\left| {h_{{s}}^{\left( 0 \right)} } \right|^2  > {{L_{{s}}^{\left( 0 \right)} {\rm{T}}} \mathord{\left/
 {\vphantom {{L_{{s}}^{\left( 0 \right)} {\rm{T}}} {\gamma ^{\left( 0 \right)} }}} \right.
 \kern-\nulldelimiterspace} {\gamma ^{\left( 0 \right)} }}} \right|L_{{s}}^{\left( 0 \right)} } \right\} \\ &\mathop  = \limits^{\left( a \right)} {1 \mathord{\left/
 {\vphantom {1 2}} \right.
 \kern-\nulldelimiterspace} 2} - ({1 \mathord{\left/
 {\vphantom {1 2}} \right.
 \kern-\nulldelimiterspace} 2}){\rm{erf}}\left( {{{\left( {\ln \left( {{{L_{{s}}^{\left( 0 \right)} {\rm{T}}} \mathord{\left/
 {\vphantom {{L_{\rm{s}}^{\left( 0 \right)} {\rm{T}}} {\gamma ^{\left( 0 \right)} }}} \right.
 \kern-\nulldelimiterspace} {\gamma ^{\left( 0 \right)} }}} \right) - \mu _s } \right)} \mathord{\left/
 {\vphantom {{\left( {\ln \left( {{{L_{\rm{s}}^{\left( 0 \right)} {\rm{T}}} \mathord{\left/
 {\vphantom {{L_{\rm{s}}^{\left( 0 \right)} {\rm{T}}} {\gamma ^{\left( 0 \right)} }}} \right.
 \kern-\nulldelimiterspace} {\gamma ^{\left( 0 \right)} }}} \right) - \mu _s } \right)} {\left( {\sqrt 2 \sigma _s } \right)}}} \right.
 \kern-\nulldelimiterspace} {\left( {\sqrt 2 \sigma _s } \right)}}} \right) \\
 & \Pr \left\{ {\left. {L_{{\rm{NLOS}}}^{\left( 0 \right)}  > L_{{\rm{LOS}}}^{\left( 0 \right)} } \right|L_{{\rm{LOS}}}^{\left( 0 \right)} } \right\}\mathop  = \limits^{\left( b \right)} \exp \left( { - \Lambda _{{\rm{NLOS}}} \left( {\left[ {0,L_{{\rm{LOS}}}^{\left( 0 \right)} } \right)} \right)} \right) \\
 & \Pr \left\{ {\left. {L_{{\rm{LOS}}}^{\left( 0 \right)}  > L_{{\rm{NLOS}}}^{\left( 0 \right)} } \right|L_{{\rm{NLOS}}}^{\left( 0 \right)} } \right\}\mathop  = \limits^{\left( c \right)} \exp \left( { - \Lambda _{{\rm{LOS}}} \left( {\left[ {0,L_{{\rm{NLOS}}}^{\left( 0 \right)} } \right)} \right)} \right) \\
 \end{split}
\end{equation}
\noindent where ${\rm{erf}}\left(  \cdot  \right)$ is the error function, (a) follows from the CDF of a Log-Normal random variable, and (b) and (c) follow from \textit{Lemma \ref{Intensity_Lemma}} and \textit{Lemma \ref{CDF_Lemma}} by using arguments similar to the computation of the intensity of $L$ and of the CDF of $L^{\left( 0 \right)}$.

The proof follows by explicitly writing the expectation with respect to $L_{{\rm{LOS}}}^{\left( 0 \right)}$ and $L_{{\rm{NLOS}}}^{\left( 0 \right)}$ in terms of their PDFs, which can be formulated, similar to (b) and (c), as $f_{L_s^{\left( 0 \right)} } \left( \xi  \right) = {{d\Pr \left\{ {L_s^{\left( 0 \right)}  < \xi } \right\}} \mathord{\left/ {\vphantom {{d\Pr \left\{ {L_s^{\left( 0 \right)}  < \xi } \right\}} {d\xi }}} \right. \kern-\nulldelimiterspace} {d\xi }} = \Lambda _s^{\left( 1 \right)} \left( {\left[ {0,\xi } \right)} \right)\exp \left( { - \Lambda _s \left( {\left[ {0,\xi } \right)} \right)} \right)$, since $\Pr \left\{ {L_s^{\left( 0 \right)}  < \xi } \right\} = \exp \left( { - \Lambda _s \left( {\left[ {0,\xi } \right)} \right)} \right)$.
\section*{Acknowledgment}
This work is supported by the European Commission under the FP7-PEOPLE MITN-CROSSFIRE project (grant 317126).
\end{document}